# The Effect of Nuclear Reaction Rates & Convective Mixing on the Evolution of a 6M$_\odot$ Star

Ghina M. Halabi

*American University of Beirut, Department of Physics, P.O. Box 11-0236, Riad El-Solh, Beirut, Lebanon, e-mail:gfm01@aub.edu.lb*

**Abstract.** We present the evolution of a 6M$_\odot$ star, of solar-like initial metallicity, and investigate the effects of key nuclear reaction rates, as well as the treatment of the convective mixing on its evolution along the Cepheid instability strip. In particular, we study the effect of recent estimates of the $^{14}$N(p,γ)$^{15}$O reaction on the formation and extension of the blue loop during core helium burning. We also investigate the effects induced on this blue loop by the adoption of non-standard convective mixing prescriptions, as well as the implications of modifying the Mixing Length Theory.

## INTRODUCTION

Stars in the mass range 5M$_\odot$ to 12M$_\odot$ develop blue loops after the onset of core helium burning, when they reach the red giant branch (RGB). This is a crossing of the Hertzsprung-Russell diagram (shortly HRD) towards higher effective temperatures and back to the RGB. Blue loops are necessary for explaining the observed non-variable yellow giants, supergiants and the δ-Cepheids in open galactic clusters ([8], [9], [12], [13]).

We focus on two factors that affect the properties of the blue loop of a 6M$_\odot$. The first is the $^{14}$N(p,γ)$^{15}$O reaction rate, since it determines the efficiency of the CNO cycle, and thus, controls the energy production during shell-hydrogen burning, which is a primary cause of the loop formation. We show that the recent evaluation of this rate (see [1]) has a strong impact on the efficiency of shell hydrogen burning (hereafter shell H-burning) which, in turn, significantly influences the behavior of the loops of intermediate mass stars (IMS).

The second factor is the treatment of convective mixing, which influences the position of the hydrogen discontinuity, and thus determines the efficiency of the H-shell burning in later stages. A non-standard mixing scheme, like envelope overshooting, is found to promote the formation and extension of blue loops of IMS; however, core overshooting during core H and He burning should be treated carefully (work under preparation). Here, we show briefly the results of mild core overshooting.

In a separate calculation, we tested the Modified Mixing Length Theory (hereafter MMLT) suggested by [3]. In their work, they argue that in the MLT, the diffusion length and blob size are the two independent parameters that determine how optically thick the blobs are. In the MMLT, a parameter $g_{ml}$ is introduced, which corresponds to the thickness of the superadiabatic region. This geometric parameter is determined using three-dimensional simulations of both adiabatic convection and stellar atmospheres without the astronomical calibration that is required in the Mixing Length Theory. We show the result in the following sections.

## STELLAR MODELS CALCULATION: EVOLUTIONARY CODE

The evolution of the 6M$_\odot$ star presented here is calculated using the stellar evolution code described by [6], [7] and references therein. The reaction rates we use are those recommended by the "JINA REACLIB" database [5]. For the $^{14}N(p,\gamma)^{15}O$ rate (hereafter referred to as the N14 rate), we use our recent evaluation as described in detail in our paper [10]. We show that this recent calculation of the rate is lower than the NACRE compilation [2] in the relevant temperature range, by approximately a factor of 2. This is also in agreement with other recent experimental determinations [1]. Convection is treated within the framework of the MLT, and we take the mixing length parameter α=2.

## RESULTS

### Effect of the $^{14}N(p,\gamma)^{15}O$ reaction rate

We present the evolution of a $6M_\odot$ star with the initial composition (X,Y,Z)=(0.7,0.28,0.02), from the zero-age main sequence till the end of core helium burning. In Fig. 1, we show the evolutionary tracks in the HRD that are obtained with the same input physics except the N14 rate, for which two different rate compilations are adopted: NACRE (left panel) and our newly calculated rate (C-rate) (right panel). The evolutionary tracks show an interesting effect of the N14 rate on the blue loop, and a detailed analysis of this result and the physics underlying it is discussed in [10]. However, here we only mention the main findings, which are summarized as follows.

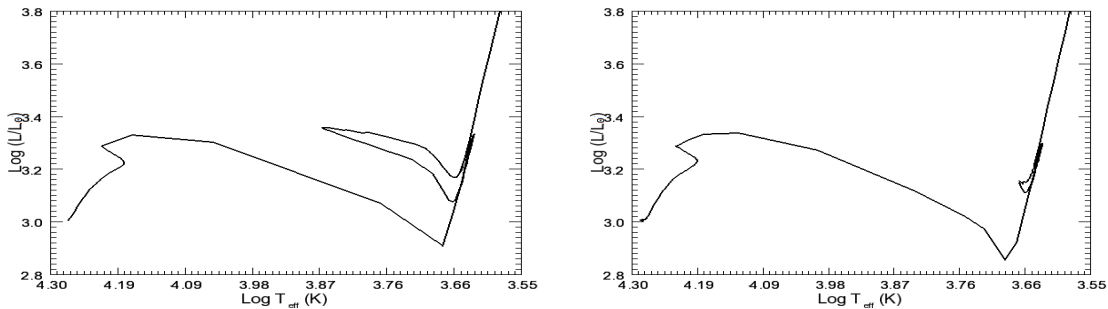

**FIGURE 1.** Evolutionary tracks in the HRD for the $6M_\odot$ evolved from the zero-age main sequence till the end of core helium burning. The only difference between the two calculations is the N14 rate: left panel: The NACRE rate is used for the $^{14}N(p,\gamma)^{15}O$ reaction, right panel: the C-rate. See text.

The $^{14}N(p,\gamma)^{15}O$ rate has a strong impact on the blue loops in the mass range comprising super-AGB stars. An appreciably reduced NACRE rate proposed on experimental ground can lead to a severe suppression of the blue loop in a $6M_\odot$ star (Fig. 1, right panel). The $^{14}N(p,\gamma)^{15}O$ rate influences the depth of the H-discontinuity during the first dredge up (FDUP) on the RGB, prior to the loop phase, and also affects the efficiency of shell H-burning during the loop phase. The less efficient shell H-burning with the weak C-rate causes an insufficient expansion of the envelope on the RGB, which results in a shallower H-profile. The combined effects of the weak C-rate and the position of the H-discontinuity cause a severely reduced loop.

### Effect of Treatment of Convection: Non-Standard Mixing & Modified Mixing Length Theory

One possibility to restore the extension of blue loops with the new N14 rate is to invoke overshooting, a form of non-standard mixing that involves the extension of full convection at the convective boundaries. Envelope overshooting places the H-discontinuity deeper in the star i.e. closer to the H-shell source. Recalculating the evolution of $6M_\odot$ with the C-rate and an exponential overshooting over a distance of $0.2H_p$ beyond the formal convective boundary is sufficient to restore the loop, as shown in Fig. 2. Invoking mild core overshooting during core H and He burning pushes the H-shell source closer towards the H-discontinuity and can also restore the blue loop as shown in Fig. 3. Observational properties of galactic and Magellanic classical Cepheids from [4] shown in the figure support the theoretical models.

In Fig. 4 we show the results of our calculation that uses the MMLT to treat convection, with the parameters $g_{ml}$ = 42 and α = 2.23 [3]. We find that this formulation has the most pronounced effect on the behavior of the convective gradient in the envelope, whereas the solution in the core, where convection is adiabatic, is barely affected. However, it doesn't have a pronounced effect on the depth of the FDUP, and therefore, the blue loop is not affected.

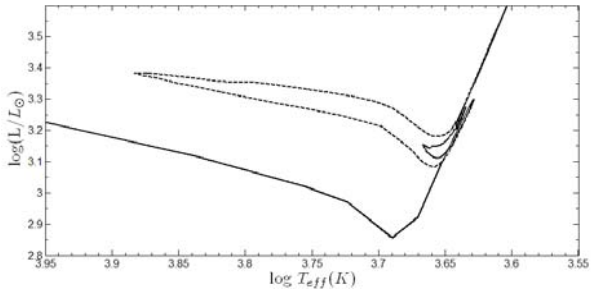

**FIGURE 2.** Solid line: the evolutionary track of a $6M_\odot$ calculated with the *C-rate* without any overshooting at the bottom of the convective envelope; the loop is almost completely suppressed. Dotted line: the same evolutionary sequence, when envelope overshooting is applied (see text). Note how the extension of the loop is restored.

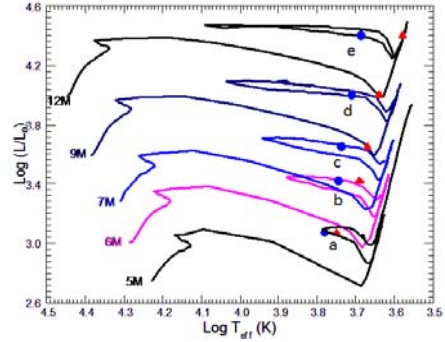

**FIGURE 3.** Evolutionary tracks of the models whose initial masses are shown, calculated with core overshooting of $0.1H_p$. We also show the luminosities and the range of the instability strip of observed galactic and Magellanic classical Cepheids (a to e) adopted from [4]. Circles mark the blue edge of the Cepheid instability strip, while triangles mark the red edge. A perfect agreement can be seen between the theoretically predicted models and the observed ones, for the five shown sequences.

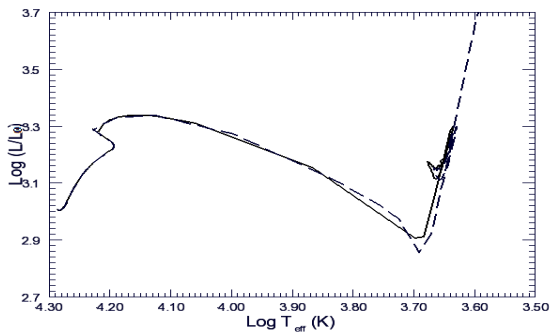

**FIGURE 4.** The HRD of the $6M_\odot$ calculated with the MMLT (solid line), and that calculated with the MLT (dashed line), both sequences without core or envelope overshooting. It is evident that the blue loop is not affected with this formulation (see text).

## CONCULSIONS

We emphasize that the $^{14}N(p,\gamma)^{15}O$ is a key reaction in the context of blue loop formation. If its rate is less efficient by about a factor of two as compared to the evaluation by NACRE, then, the blue loops can be recovered by introducing moderate envelope and core overshooting. Moreover, we tested the effects of the suggested Modified Mixing Length Theory on the extension of blue loops. We found that this modification has no effect that can correct for the suppression of the blue loop of the $6M_\odot$ model.

## ACKNOWLEDGMENTS

I wish to thank the organisers of the 7[th] European Summer School of Nuclear Astrophysics, and in particular Prof. Claudio Spitaleri and Prof. Maurizio Busso, for the very generous support that made my participation in the conference possible.

## REFERENCES

1. Adelberger, E. G., et al., *Rev. Mod. Phys*. 83, 195A (2011)
2. Angulo, C., et al., *Nucl. Phys*. A656, 3A (1999)
3. Arnett, D., Meakin, C., Young, P. A., ApJ 710, 1619 (2010)
4. Criscienzo, M. Di, Marconi, M., Musella, I., Cignoni, M. & Ripepi, V., MNRAS sts023, 36(2012)
5. Cyburt et al., *ApJS* 189, 240 (2010)
6. El Eid M. F., The L. .S., Meyer B., *ApJ* 611, 452 (2004)
7. El Eid M. F., The L. .S., Meyer B., *SSRv* 147, 1E (2009)
8. Evans, N. R., *AJ* 105, 1956 (1993)
9. Fernie, J. D., Evans, N. R., Beattie, B., & Seager, S*., IBVS* 4148, 1F (1995)
10. Halabi, G. M., El Eid, M. F. & Champagne A., ApJ 761, 10 (2012)
11. Imbriani et al., *Eur. Phy*. J. A 25, 455 (2005)
12. Mermilliod, J. C*., A&AS* 44, 467 (1981)
13. Schmidt, E. G*., ApJ* 287, 261 (1984)